\documentclass[
 reprint,
bibnotes,
 amsmath,amssymb,
 aps,
prb,
floatfix,
]{revtex4-2}
\usepackage{graphicx}
\usepackage{dcolumn}
\usepackage{bm}

\usepackage{hyperref}
\hypersetup{
  colorlinks=true,
  linkcolor=blue,
  citecolor=blue,
  urlcolor=blue,
}
\usepackage[mathlines]{lineno}
\usepackage{mathtools}
\usepackage{leftidx}
\usepackage{physics}
\usepackage{tikz}
\usepackage{pgfplots}
\pgfplotsset{compat=1.16}
\usetikzlibrary{intersections,calc,arrows,arrows.meta,shapes,bending,positioning,decorations.markings,backgrounds}
\usepgfplotslibrary{patchplots}
\usepgfplotslibrary{fillbetween}
\pgfplotsset{%
    layers/standard/.define layer set={%
        background,axis background,axis grid,axis ticks,axis lines,axis tick labels,pre main,main,axis descriptions,axis foreground%
    }{
        grid style={/pgfplots/on layer=axis grid},%
        tick style={/pgfplots/on layer=axis ticks},%
        axis line style={/pgfplots/on layer=axis lines},%
        label style={/pgfplots/on layer=axis descriptions},%
        legend style={/pgfplots/on layer=axis descriptions},%
        title style={/pgfplots/on layer=axis descriptions},%
        colorbar style={/pgfplots/on layer=axis descriptions},%
        ticklabel style={/pgfplots/on layer=axis tick labels},%
        axis background@ style={/pgfplots/on layer=axis background},%
        3d box foreground style={/pgfplots/on layer=axis foreground},%
    },
}

\usepackage{color}
\definecolor{darkcyan}{rgb}{0.0, 0.55, 0.55}

\newcommand*{\halfway}{0.5*\pgfdecoratedpathlength+9.75pt}

\begin{document}

\preprint{APS/123-QED}

\title{Tensor network simulations for nonorientable surfaces}
\author{Haruki Shimizu}
\email{shimizu-haruki@issp.u-tokyo.ac.jp}
\author{Atsushi Ueda}
\affiliation{%
 Institute for Solid State Physics, University of Tokyo, Kashiwa 277-8581, Japan
}


\date{\today}

\begin{abstract}
In this study, we explore the geometric construction of the Klein bottle and the real projective plane ($\mathrm{RP}^2$) within the framework of tensor networks, focusing on the implementation of crosscap and rainbow boundaries. Previous investigations have applied boundary matrix product state techniques to study these boundaries. We introduce an approach that incorporates such boundaries into the tensor renormalization group methodology, facilitated by an efficient representation of a spatial reflection operator. This advancement enables us to compute the crosscap and rainbow free energy terms and the one-point function on $\mathrm{RP}^2$ with enhanced efficiency and for larger system sizes. Additionally, our method is capable of calculating the partition function under isotropic conditions of space and imaginary time. The versatility of this approach is further underscored by its applicability to constructing other (non)orientable surfaces of higher genus.
\end{abstract}

\maketitle


\section{\label{sec:intro}Introduction}
Classifying the phases of matter represents a fundamental challenge and goal within the field of physics. This classification enables the tailored design of models to exhibit desired phenomena, navigating the intricate landscape of physical properties. However, the complexity inherent in realistic models presents significant challenges. To address these, the concept of universality in critical phenomena occurring at the phase boundaries becomes instrumental. Universal quantities, such as critical exponents, offer insights into the mechanisms underpinning phase transitions. These insights are invaluable, as the fundamental processes of phase transitions are often not apparent through the microscopic details of the systems involved.

The utility of universality is underscored in the study of critical phenomena within both one-dimensional quantum systems and two-dimensional classical systems, where conformal field theory (CFT) serves to describe these universal properties. Recent advancements have further highlighted the importance of CFT-derived universal quantities in analyzing edge states within (2+1)-dimensional symmetry-protected topological phases, making the acquisition of CFT data crucial for understanding the underlying mechanisms of phase transitions~\cite{PhysRevB.85.245132,PhysRevB.88.075125,PhysRevB.96.125105}.

An intriguing development in this context is that universal quantities can be discerned through the partition function on nonorientable surfaces. This approach indicates that a spatial reflection introduces a universal term to the partition function. In particular, the ratio of the partition function on the Klein bottle to that on the torus is given by a universal ratio~\cite{Fioravanti_1994,tu_universal_2017}:
\begin{align}
\frac{Z^\mathcal{K}(\frac{\beta}{2},2L)}{Z^\mathcal{T}(\beta,L)}\simeq g, \label{eq:zk-scaling}
\end{align}
when the system length is significantly larger than its inverse temperature $L\gg\beta$. Here, $g$ is a universal constant unique to its universality class. More precisely, $g$ is $\sum_a\frac{d_a}{\mathcal{D}}$ for diagonal CFTs, with $d_a$ being the quantum dimension of the primary field $a$, and $\mathcal{D}$ representing the total quantum dimension. As $g$ takes a unique value depending on the criticality, it can be used as a tool to detect phase transitions ($g=1$ for trivially gapped phases). This theory has been corroborated through numerical studies, including Monte Carlo and tensor-network simulations employing boundary matrix product state (BMPS) techniques~\cite{tang_universal_2017,PhysRevLett.86.2134,PhysRevE.63.026107,tang_klein_2019,wang_topological_2018,li_critical_2020,chen_conformal_2017,li_tensor_2021}.

This concept was further expanded to include the nonorientable surface $\mathrm{RP}^2$, leading to the discovery of another universal ratio \cite{Maloney_2016,wang_topological_2018}:
\begin{align}
\frac{Z^{\mathrm{RP}^2}(\frac{\beta}{2}, 2L)}{Z^\mathcal{T}(\beta,L)} \sim \beta^{c/4}, \label{eq:zrp2-scaling}
\end{align}
where $c$ denotes the central charge, a critical universal parameter within CFT. The identification of these universal ratios is pivotal for elucidating the universal critical and topological characteristics of systems. Yet, direct simulation of geometrically twisted boundaries, crucial for the realization of both the Klein bottle and $\mathrm{RP}^2$, presents considerable challenges. As a workaround, previous efforts have resorted to approximations, such as computing crosscap and rainbow states via BMPS techniques, applicable predominantly in the anisotropic limit where $L \gg \beta$. Given the foundational role of $\mathrm{RP}^2$ in constructing generalized nonorientable surfaces, its accurate realization across a broader range of cases remains a critical objective.

Addressing this gap, we introduce tensor-network strategies for simulating the Klein bottle and $\mathrm{RP}^2$ configurations directly. By integrating the higher-order tensor renormalization group (HOTRG)~\cite{Xie_2012} with spatial reflections, we enable the computation of $Z^\mathcal{K}$ and $Z^{\mathrm{RP}^2}$ for extensive system sizes, thereby facilitating a deeper understanding of universal critical and topological features.

In addition to this, we demonstrate that HOTRG can be utilized to extract another piece of universal data. The partition function of CFT on a torus is determined by its operator content and operator product expansion coefficients $C_{ijk}$. Prior research has illustrated the effectiveness of both tensor renormalization group (TRG) and tensor network renormalization (TNR) methods in efficiently retrieving these CFT data~\cite{PhysRevLett.99.120601,PhysRevB.80.155131,PhysRevLett.115.180405,PhysRevB.95.045117,PhysRevLett.118.110504,PhysRevLett.118.250602,PhysRevB.97.045111,PhysRevResearch.6.043102,Ueda_2023,PhysRevLett.116.040401,PhysRevResearch.4.023159,ueda2023fixedpoint}. For nonorientable surfaces, however, a different type of universal data is required—the one-point function of the primary operators~\cite{Fioravanti_1994}:
\begin{align}
\langle\phi_k\rangle_{\mathrm{RP}^2} = \frac{\Gamma_k}{(1+z\bar{z})^{2h}},\label{eq:Gammak}
\end{align}
with $\mathrm{RP}^2$ conceptualized as the complex plane $z$, under the involution $I(z) = \frac{-1}{\bar{z}}$ and $h$ being a conformal weight. This conceptualization stems from viewing $\mathrm{RP}^2$ as a projective space in which antipodal points on a Riemann sphere of radius $\frac{1}{2}$ are identified. Our findings indicate that the parameter $\Gamma_k$, pivotal for understanding the behavior of physical quantities on nonorientable Riemann surfaces, can be directly calculated using both TRG and TNR approaches.

This paper is organized as follows: Initially, we provide an overview of tensor network methodologies and the concept of CFT on the Klein bottle and $\mathrm{RP}^2$ in Sec.~\ref{sec:method}. Subsequently, the numerical outcomes derived from our algorithm are showcased in Sec.~\ref{sec:results}. Finally, we explore the potential implications and applicability of our findings in Sec.~\ref{sec:discussion-summary}.

Our numerical calculations were performed using ITensors~\cite{10.21468/SciPostPhysCodeb.4,*10.21468/SciPostPhysCodeb.4-r0.3}. Sample code to reproduce the figures in Sec.~\ref{sec:results} can be found in Ref.~\footnote{GitHub, \url{https://github.com/elle-et-noire/HOTRG-nonorientablesurface}}.

\section{\label{sec:method}Method}
In this study, we focus on statistical models defined on a square lattice with nearest neighbor interactions.
Although our demonstrations primarily utilize the HOTRG framework, the methodologies we introduce are versatile enough to be applied across a range of TRG and TNR algorithms.
\subsection{Review on HOTRG}
In statistical mechanics, the probability of a specific configuration is captured by its Boltzmann weight. The aggregate of these weights across all possible configurations yields the partition function, a cornerstone for calculating various physical properties. Within the framework of tensor network representations, this comprehensive summation process is facilitated by tensor contractions. Contracting tensors effectively sums over the indices being contracted, which correspond to the degrees of freedom in the system, such as a spin in lattice models. 

Let $T^{(0)}=T$ be an initial tensor of HOTRG which represents the local Boltzmann weights of the partition function.
The torus partition function is obtained by contracting all the tensors in the system with bonds connecting the nearest-neighboring sites:
\begin{equation*}
    Z = \sum_{\substack{\{x_i\},\{y_j\} \\\quad\in \{1,\dots,D\}^L}}\prod_{1\le i,j\le L} T_{x_i x_{i+1} y_j y_{j+1}},
\end{equation*}
where $x_{L+1} = x_1$ and $y_{L+1} = y_1$.

At the $n$th step of coarse graining, first we contract the previously obtained tensor $T^{(n-1)}$ in the vertical direction:
\begin{equation}
    M^{(n)}_{x_1x_2x_1'x_2'yy'}=\sum_i T^{(n-1)}_{x_1x_1'yi}T^{(n-1)}_{x_2x_2'iy'}. \label{eq:tt}
\end{equation}
Then we obtain the isometry by singular value decomposition (SVD):
\begin{equation*}
    \sum_{xx'yy'}M^{(n)}_{x_1x_2xx'yy'}M^{(n)}_{x_1'x_2'xx'yy'} \approx \sum_{ii'} U^{(n)}_{ix_1x_2} \Lambda^{(n)}_{ii'} U^{(n)}_{i'x_1'x_2'}
\end{equation*}
where the bond dimension of $i$ and $i'$ are truncated up to $\chi$. Contracting the isometry $U^{(n)}$, we get a vertically renormalized tensor:
\begin{equation*}
    W^{(n)}_{xx'yy'}=\ \smashoperator{\sum_{x_1x_1'y_1y_1'}}\ U^{(n)}_{xx_1x_2}M^{(n)}_{x_1x_2x_1'x_2'yy'}U^{(n)}_{x'x_1'x_2'}.
\end{equation*}

Then, we repeat the same procedure in the horizontal direction:
\begin{equation*}
    T^{(n)}_{xx'yy'}=\ \smashoperator{\sum_{iy_1y_1'y_2y_2'}}\ V^{(n)}_{yy_1y_2}W^{(n)}_{xiy_1y_1'}W^{(n)}_{ix'y_2y_2'}V^{(n)}_{y'y_1'y_2'}
\end{equation*}
with an appropriate isometry $V^{(n)}$ obtained by SVD. Thus, the height and the width of the system renormalized into $T^{(n)}$ are equal.

\subsection{Renormalization of a spatial reflection operator}
\begin{figure}[tb]
    \centering
    \includegraphics[width=86mm]{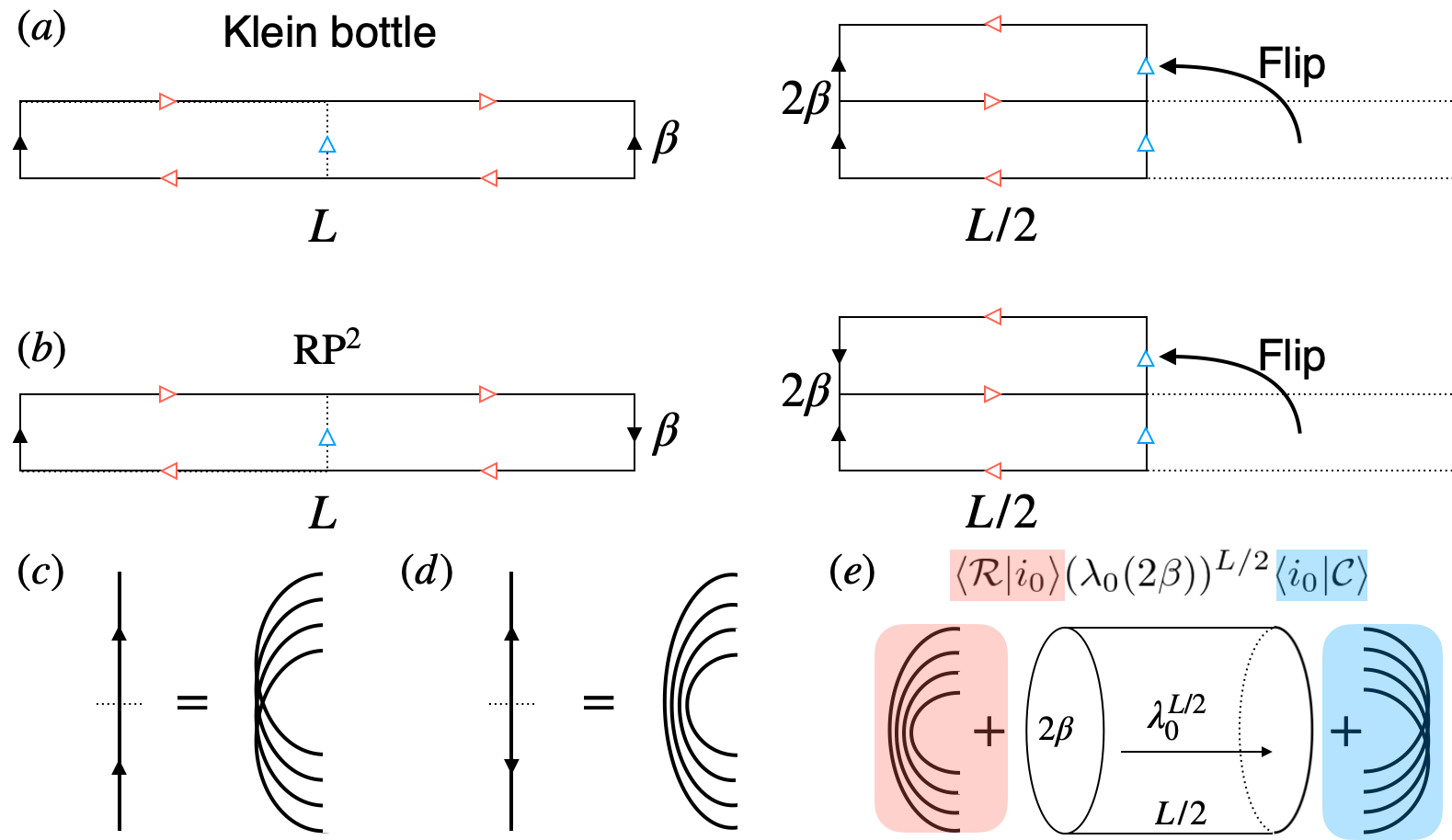}
    \caption{Illustrations of the partition functions on (a) the Klein bottle and (b) $\mathrm{RP}^2$. These partition functions can be cut in the middle as shown in the blue arrows(left panel), and then sewn again after flipping the latter half in $x$-direction. The resulting boundaries are crosscap states for the Klein bottle and rainbow and crosscap states for $\mathrm{RP}^2$. The schematic pictures of the crosscap and rainbow boundaries are shown in (c) and (d), respectively. (e) In the limit $L\gg\beta\gg1$, the bulk part of the partition function $Z^{\mathrm{RP}^2}(\beta,L)$ is dominated by the leading eigenvector $|i_0\rangle$ of the transfer matrix on a cylinder. The boundary contributions are then represented as the overlaps with $|i_0\rangle$ as denoted as $\langle\mathcal{R}|i_0\rangle$ and $\langle i_0|\mathcal{C}\rangle$.}
    \label{fig_sew}
\end{figure}
Consider a system characterized by a width $L$ along the $x$ direction and a height $\beta$ along the $y$ direction, adhering to the thermodynamic limit $L \gg \beta \gg 1$. To compute the partition functions for the Klein bottle and $\mathrm{RP}^2$ within a tensor network framework, we employ the cut-and-sew method as described in Refs.~\cite{Fioravanti_1994, wang_topological_2018}. We briefly review these concepts for clarity. 

The partition function on the Klein bottle is periodic in the spatial direction and undergoes a geometric twist in the $\beta$-direction, as depicted in Fig.~\ref{fig_sew}(a). To evaluate this quantity, we bisect the partition function medially, then overlay the right half atop the left, subsequent to a spatial inversion. This process is visualized in the right panel of Fig.~\ref{fig_sew}(a). The resultant partition function has a width of $L/2$ and a height of $2\beta$. Notably, the $\beta$-direction retains its periodicity, whereas the boundaries at $x=0$ and $L/2$ are no longer periodic. Instead, they adopt crosscap boundary conditions, as illustrated in $(c)$.

A similar ``cut and sew" process applied to $\mathrm{RP}^2$ results in the imposition of rainbow and crosscap boundaries at $x=0$ and $L/2$, respectively, as depicted in Fig.~\ref{fig_sew}(b). The rainbow boundary is further illustrated in Fig.~\ref{fig_sew}(d). Notably, the $\beta$ direction maintains periodicity in $\mathrm{RP}^2$ as well. This periodicity is instrumental in facilitating the computation of partition functions on both the Klein bottle and $\mathrm{RP}^2$.

Let $M_\beta$ be a transfer matrix in the $x$ direction, which has height $\beta$. In the limit $L\gg\beta\gg1$, the partition function $Z(\beta,L) = \mel*{\mathcal{B}_L}{M_\beta^L}{\mathcal{B}_R}$ is approximated using the largest eigenvector and its eigenvalue, respectively denoted as $|i_0\rangle_\beta$ and $\lambda_0(\beta)$:
\begin{equation*}
    Z(\beta,L) \stackrel{L\gg \beta}{\approx} \braket{\mathcal{B}_L}{i_0}_\beta[\lambda_{0}(\beta)]^L\leftidx{_\beta}{\hspace{-0.1em}\braket{i_0}{\mathcal{B}_R}}{}.
\end{equation*}

In the case of the torus, the Klein bottle and $\mathrm{RP}^2$, the corresponding partition functions are given as
\begin{align}
    Z^\mathcal{T}(\beta,L) &\approx [\lambda_{0}(\beta)]^L \label{eq:z-torus},\\
    Z^\mathcal{K}(\beta,L) &\approx \braket{\mathcal{C}}{i_0}_{2\beta}[\lambda_0(2\beta)]^{L/2}{\phantom{\lvert}}_{2\beta}\hspace{-0.2em}\braket{i_0}{\mathcal{C}} \label{eq:z-klein},\\
    Z^{\mathrm{RP}^2}(\beta,L) &\approx \braket{\mathcal{R}}{i_0}_{2\beta}[\lambda_{0}(2\beta)]^{L/2}{\phantom{\lvert}}_{2\beta}\hspace{-0.2em}\braket{i_0}{\mathcal{C}}\label{eq:RP2_assymptotic},
\end{align}
where $\ket{\mathcal{C}}$ ($\ket{\mathcal{R}}$) represents a crosscap (rainbow) state. Equation~\eqref{eq:RP2_assymptotic} is pictorially shown in Fig.~\ref{fig_sew} $(e)$. 

Given Eqs.~(\ref{eq:z-torus})--(\ref{eq:RP2_assymptotic}), $Z^{\mathcal{K}}$ and $Z^{\mathrm{RP}^2}$ contain extra terms arising from the boundaries compared to $Z^{\mathcal{T}}$ as follows:
\begin{align*}
     \ln \dfrac{Z^{\mathcal{K}}(\frac{\beta}{2}, 2L)}{Z^{\mathcal{T}}(\beta, L)} &= 2\ln \abs{\braket{\mathcal{C}}{i_0}}, \\
    \ln \dfrac{Z^{\mathrm{RP}^2}(\frac{\beta}{2}, 2L)}{Z^{\mathcal{T}}(\beta, L)} &= \ln\abs{\braket{\mathcal{C}}{i_0}} + \ln\abs*{\braket{\mathcal{R}}{i_0}_\beta}
\end{align*}
These boundary free energies, defined as $F_\mathcal{C}\coloneqq\ln\abs{\braket{\mathcal{C}}{i_0}}$ and $F_\mathcal{R}\coloneqq\ln\abs*{\braket{\mathcal{R}}{i_0}_\beta}$, are respectively referred to as crosscap and rainbow free energy.
In CFT, $F_\mathcal{C}$ and $F_\mathcal{R}$ are expressed with universal terms as~\cite{tang_universal_2017,wang_topological_2018}
\begin{align}
    F_\mathcal{C} &= \frac{1}{2}\ln g,\\
    F_\mathcal{R} &= \frac{c}{4}\ln \beta + b,
\end{align}
where $c$ and $b$ are the central charge and a nonuniversal constant. Thus, computing the crosscap and rainbow free energy allows us to directly read out the universal data.

To compute $F_\mathcal{C}$ or $F_\mathcal{R}$ via tensor network techniques, consider the scenario at the $n$th step of HOTRG, specifically when $\beta = 2^n$. Here, the matrix $M_\beta$ is essentially derived from $T^{(n)}$ by contracting its indices along the $y$ direction. The leading eigenvector of $M_\beta$, denoted as $\ket{i_0}_\beta$, is required to possess two tensor indices to facilitate contraction with the states $\ket{\mathcal{C}}$ or $\ket{\mathcal{R}}$, each of which also manifests in a dual-index tensor format as elucidated subsequently. This necessitates the vertical duplication of $T^{(n-1)}$, leading to the following formulation:
\begin{equation*}
    (M_{\beta})_{(x_1x_2)(x_1'x_2')} = \sum_y M^{(n)}_{x_1x_2x_1'x_2'yy}\ {},
\end{equation*}
where $(M_{\beta})_{(x_1x_2)(x_1'x_2')}$ represents the transfer matrix along the $x$ direction. In this context, $(x_1x_2)$ and $(x_1'x_2')$ serve as the composite matrix indices, with the notation $(ab)$ signifying the combining of indices $a$ and $b$.

Let ${L}_{x_1x_2}^{(n)}$ be the leading eigenvector of $M_\beta$:
\begin{equation*}
    \sum_{x_1x_2}L_{x_1x_2}^{(n)} (M_\beta)_{(x_1x_2)(x_1'x_2')} = \lambda_{0}(\beta) L_{x_1'x_2'}^{(n)}.
\end{equation*}
Then we can assume $L^{(n)}_{x_1x_2}$ as a tensor representation of $\ket{i_0}_\beta$ with two indices.

\begin{figure}[tb]
    \centering
    \scalebox{0.5}{\begin{tikzpicture}[line width = 1.6pt,decoration={markings,mark=at position \halfway with {\arrow[black]{Stealth[scale=1.5]}}}]
  \begin{scope}
    \foreach \y in {3,1,...,-3} {
      \filldraw[fill=lime] (0, \y + 0.7)--(0, \y - 0.7)--(-0.448, \y)--cycle;
      \draw[postaction={decorate}] (0, \y + 0.7)--(0, \y - 0.7);
      \draw (0, \y + 0.4666)--(0.5, \y + 0.4666);
      \draw (0, \y - 0.4666)--(0.5, \y - 0.4666);
      \draw (-0.448, \y)--(-1.07, \y);
    }

    \foreach \y in {2,-2} {
      \filldraw[fill=lime] (-1.07, \y + 1.5)--(-1.07, \y - 1.5)--(-1.07 - 0.96, \y)--cycle;
      \draw[postaction={decorate}] (-1.07, \y + 1.5)--(-1.07, \y - 1.5);
    }
    \draw (-1.07 - 0.96, 2) to [out = 180, in = 180] (-1.07 - 0.96, -2);
    \draw (1.4, 0)node{{\Huge$\approx$}};
  \end{scope}

  \begin{scope}[xshift = 4.1cm, every node/.style={sloped,allow upside down}]
    \foreach \y in {3,1,...,-3} {
      \filldraw[fill=lime] (0, \y + 0.7)--(0, \y - 0.7)--(-0.448, \y)--cycle;
      \draw[postaction={decorate}] (0, \y + 0.7)--(0, \y - 0.7);
      \draw (0, \y + 0.4666)--(0.5, \y + 0.4666);
      \draw (0, \y - 0.4666)--(0.5, \y - 0.4666);
    }
    \foreach \y in {1,-1} \draw (-0.448, \y + 2) to [out = 180, in = 180] (-0.448, \y - 2);
    \draw (1.4, 0)node{{\Huge$\approx$}};
  \end{scope}

  \begin{scope}[xshift = 8cm]
  \foreach \y in {3,1,...,-3} {
    \draw (0, \y + 0.4666)--(0.5, \y + 0.4666);
    \draw (0, \y - 0.4666)--(0.5, \y - 0.4666);
  }
  \foreach \y in {1,-1} {
    \draw (0, \y + 2 + 0.4666) to [out = 180, in = 180] (0, \y - 2 + 0.4666);
    \draw (0, \y + 2 - 0.4666) to [out = 180, in = 180] (0, \y - 2 - 0.4666);
  }
  \end{scope}
\end{tikzpicture}}
    \caption{Tensor network representations of a crosscap boundary condition.
    The six triangles are the left-hand side isometries to renormalize the bulk tensor $T^{(0)}$ into $T^{(2)}$,
    with four smaller ones being $U^{(1)}$ and two larger ones $U^{(2)}$.
    The arrows indicate the order of two indices before amalgamation and truncation,
    which correspond to the order of $x_1$ and $x_2$ in $U^{(n)}_{xx_1x_2}$.}
    \label{fig:crosscap}
\end{figure}
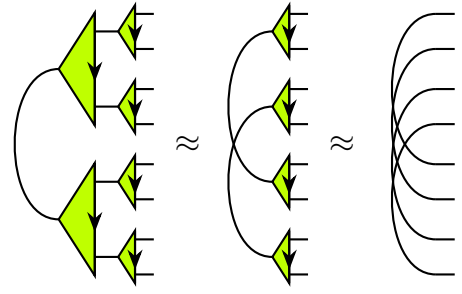

\subsubsection{Crosscap free energy term\label{sec:crosscap}}

To construct a crosscap within the coarse-grained tensor framework, we proceed by contracting the indices $x_1$ and $x_2$ of $L^{(n)}_{x_1x_2}$, as illustrated in Fig.~\ref{fig:crosscap}. This operation effectively encapsulates the crosscap structure at the $n$th level of coarse graining. The crosscap free energy ${F}_\mathcal{C}^{(n)}$ can subsequently be computed as follows:
\begin{equation}
{F}_\mathcal{C}^{(n)} = \ln\qty\Bigg(\sum_{x}{L}^{(n)}_{xx}).\label{eq:crosscap}
\end{equation}
This equation provides a direct method for evaluating the free energy associated with the crosscap configuration, leveraging the simplicity of tensor contractions to elucidate the energetic contributions of topologically non-trivial structures within the tensor network formalism.

\subsubsection{Rainbow free energy term}
\begin{figure}[tb]
    \centering
    \scalebox{0.5}{\begin{tikzpicture}[line width = 1.6pt,decoration={markings,mark=at position \halfway with {\arrow[black]{Stealth[scale=1.5]}}}]
  \begin{scope}
    \foreach \y in {3,1,...,-3} {
      \filldraw[fill=lime] (0, \y + 0.7)--(0, \y - 0.7)--(-0.448, \y)--cycle;
      \draw[postaction={decorate}] (0, \y + 0.7)--(0, \y - 0.7);
      \draw (0, \y + 0.4666)--(0.5, \y + 0.4666);
      \draw (0, \y - 0.4666)--(0.5, \y - 0.4666);
      \draw (-0.448, \y)--(-1.07, \y);
    }

    \foreach \y in {2,-2} {
      \filldraw[fill=lime] (-1.07, \y + 1.5)--(-1.07, \y - 1.5)--(-1.07 - 0.96, \y)--cycle;
      \draw[postaction={decorate}] (-1.07, \y + 1.5)--(-1.07, \y - 1.5);
    }
    \node[fill=orange!75,draw,circle,minimum size=30pt] (o3) at (-1.07 - 0.96 - 1, 0){$O$};
    \draw (-1.07 - 0.96, 2) to [out = 180, in = 90] (o3) to [out = -90, in = 180] (-1.07 - 0.96, -2);
    \draw (1.4, 0)node{{\Huge$\approx$}};
  \end{scope}

  \begin{scope}[xshift = 4.7cm, every node/.style={sloped,allow upside down}]
    \foreach \y in {3,1,...,-3} {
      \filldraw[fill=lime] (0, \y + 0.7)--(0, \y - 0.7)--(-0.448, \y)--cycle;
      \draw[postaction={decorate}] (0, \y + 0.7)--(0, \y - 0.7);
      \draw (0, \y + 0.4666)--(0.5, \y + 0.4666);
      \draw (0, \y - 0.4666)--(0.5, \y - 0.4666);
    }

    \node[fill=orange!75,draw,circle] (o1) at (-0.448 - 0.5, 0){$O$};
    \node[fill=orange!75,draw,circle] (o2) at (-0.448 - 1.5, 0){$O$};
    \foreach \y/\o in {1/o1, 3/o2} {
      \draw (-0.448, \y) to [out = 180, in = 90] (\o) to [out = -90, in = 180] (-0.448, -\y);
    }
    \draw (1.4, 0)node{{\Huge$\approx$}};
  \end{scope}

  \begin{scope}[xshift = 9.4cm]
    \foreach \y in {3,1,...,-3} \draw (0, \y + 0.4666)--(0.5, \y + 0.4666);
    \foreach \y in {3,1,...,-3} \draw (0, \y - 0.4666)--(0.5, \y - 0.4666);
    \foreach \y in {1,-1} \draw (0, \y + 2 + 0.4666) to [out = 180, in = 180] (0, -\y - 2 - 0.4666);
    \foreach \y in {1,-1} \draw (0, \y - 2 + 0.4666) to [out = 180, in = 180] (0, -\y + 2 - 0.4666);
  \end{scope}
\end{tikzpicture}}
    \caption{Tensor network representations of a rainbow boundary condition. With the spatial reflection operator $O^{(n)}$ in between the isometries, the lower half of the indices are in the reverse order. }
    \label{fig:rainbow-sro}
\end{figure}
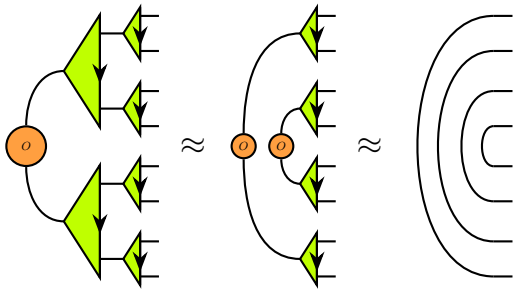

\begin{figure}[tb]
    \centering
    \scalebox{0.5}{\begin{tikzpicture}[line width = 1.6pt,decoration={markings,mark=at position \halfway with {\arrow[black]{Stealth[scale=1.5]}}}]
  \begin{scope}
    \foreach \y in {3,1,...,-3} {
      \filldraw[fill=lime] (0, \y + 0.7)--(0, \y - 0.7)--(-0.448, \y)--cycle;
      \pgfmathtruncatemacro\result{2 * mod((\y + 3) / 2, 2) - 1}
      \draw[postaction={decorate}] (0, \y + 0.7 * \result)--(0, \y - 0.7 * \result);
      \draw (0, \y + 0.4666)--(0.5, \y + 0.4666);
      \draw (0, \y - 0.4666)--(0.5, \y - 0.4666);
      \draw (-0.448, \y)--(-1.07, \y);
    }

    \foreach \y in {2,-2} {
      \filldraw[fill=lime] (-1.07, \y + 1.5)--(-1.07, \y - 1.5)--(-1.07 - 0.96, \y)--cycle;
      \pgfmathtruncatemacro\result{sign(\y)}
      \draw[postaction={decorate}] (-1.07, \y + 1.5 * \result)--(-1.07, \y - 1.5 * \result);
    }
    \draw (-1.07 - 0.96, 2) to [out = 180, in = 180] (-1.07 - 0.96, -2);
    \draw (1.4, 0)node{{\Huge$\approx$}};
  \end{scope}

  \begin{scope}[xshift = 4.7cm, every node/.style={sloped,allow upside down}]
    \foreach \y in {3,1,...,-3} {
      \filldraw[fill=lime] (0, \y + 0.7)--(0, \y - 0.7)--(-0.448, \y)--cycle;
      \pgfmathtruncatemacro\result{2 * mod((\y + 3) / 2, 2) - 1}
      \draw[postaction={decorate}] (0, \y + 0.7 * \result)--(0, \y - 0.7 * \result);
      \draw (0, \y + 0.4666)--(0.5, \y + 0.4666);
      \draw (0, \y - 0.4666)--(0.5, \y - 0.4666);
    }

    \foreach \y/\o in {1/o1, 3/o2} {
      \draw (-0.448, \y) to [out = 180, in = 180] (-0.448, -\y);
    }
    \draw (1.4, 0)node{{\Huge$\approx$}};
  \end{scope}

  \begin{scope}[xshift = 9.4cm]
    \foreach \y in {3,1,...,-3} \draw (0, \y + 0.4666)--(0.5, \y + 0.4666);
    \foreach \y in {3,1,...,-3} \draw (0, \y - 0.4666)--(0.5, \y - 0.4666);
    \foreach \y in {1,-1} \draw (0, \y + 2 + 0.4666) to [out = 180, in = 180] (0, -\y - 2 - 0.4666);
    \foreach \y in {1,-1} \draw (0, \y - 2 + 0.4666) to [out = 180, in = 180] (0, -\y + 2 - 0.4666);
  \end{scope}
\end{tikzpicture}}
    \caption{Alternative way to construct a rainbow boundary condition in the tensor network. Instead of contracting a renormalized spatial reflection operator, we just reflect the vertically copied bulk tensor $T^{(n)}$ in the vertical direction, which means that the included isometries are also reflected.}
    \label{fig:rainbow-reflect}
\end{figure}
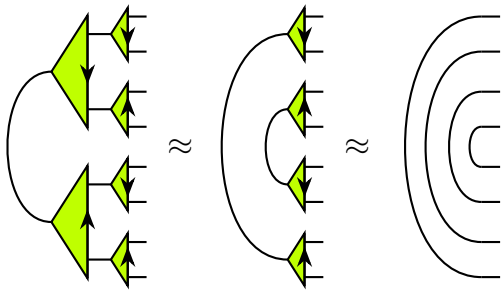

In order to construct a rainbow state, we renormalize a spatial reflection operator which inverts the direction of an edge of the bulk tensor $T^{(n)}$.

Let the initial spatial reflection operator be a unit matrix:
\begin{equation*}
    O^{(0)}_{ab} = \delta_{ab}.
\end{equation*}
At the $n$-th step, we renormalize the spatial reflection operator reusing the isometry used to renormalize $T^{(n)}$:
\begin{equation*}
    O^{(n)}_{ab} = \sum_{ii'jj'}U^{(n)}_{aij}O^{(n-1)}_{ii'}O^{(n-1)}_{jj'}U^{(n)}_{bj'i'}.
\end{equation*}
This can be graphically represented as follows:
\begin{equation*}
    \scalebox{0.6}{\begin{tikzpicture}[line width = 1.6pt,decoration={markings,mark=at position \halfway with {\arrow[black]{Stealth[scale=1.5]}}}]
  \coordinate (y1) at (-1, 1);
  \coordinate (y2) at (-1, -1);
  \coordinate (y3) at (1, 1);
  \coordinate (y4) at (1, -1);
  \coordinate (p1) at (-1.64, 0);
  \coordinate (p2) at (1.64, 0);
  \filldraw[fill=lime] (y1)--(y2)--(p1)--cycle;
  \filldraw[fill=lime] (y3)--(y4)--(p2)--cycle;
  \coordinate (a) at (1, 0.6666);
  \coordinate (b) at (-1, -0.6666);
  \coordinate (c) at (1, -0.6666);
  \coordinate (d) at (-1, 0.6666);
  \node[fill=orange!75,draw,circle] (o1) at (0.3, 0.6666) {$O$};
  \node[fill=orange!75,draw,circle] (o2) at (0.3, -0.6666) {$O$};
  \draw (a)--(o1) to [out = 190, in = 35] (b);
  \draw (c)--(o2) to [out = 170, in = -35] (d);
  \draw [postaction={decorate}](y1)--(y2);
  \draw [postaction={decorate}](y3)--(y4);
  \draw (p2)--(2.14, 0);
  \draw (p1)--(-2.14, 0);
  \draw [->, >=stealth] (2.64, 0)--(3.5, 0);
  \node[fill=orange!75,draw,circle,minimum size=30pt] (o3) at (5, 0){$O$};
  \draw (4, 0)--(o3)--(6, 0);
\end{tikzpicture}}.
\end{equation*}
Using the spatial reflection operator and the leading eigenvector $L^{(n)}_{x_1x_2}$, we can construct a rainbow boundary in the coarse-grained tensor and obtain the $n$th rainbow term, $F_\mathcal{R}^{(n)}$, as follows~\footnote{Note that the order $n$ of $O^{n}$ is crucial. The indices $x_1$ and $x_2$ of $L^{(n)}_{x_1x_2}$ are coming from the left-hand side index of $T^{(n-1)}$, which is originally the truncated index of $U^{(n-1)}$. Because $O^{(n-1)}$ also has indices of $U^{(n-1)}$, $L^{(n)}$ has to be contracted with $O^{(n-1)}$.}:
\begin{equation}
    {F}_\mathcal{R}^{(n)} = \ln\qty\Bigg(\,\sum_{x_1x_2}{L}^{(n)}_{x_1x_2}O^{(n-1)}_{x_1x_2}) \label{eq:fr-sro}.
\end{equation}
This process is depicted in Fig.~\ref{fig:rainbow-sro}.

An alternative approach to constructing a rainbow state involves inverting the direction of the copied tensor at the vertical contraction step in Eq.~\eqref{eq:tt}, which effectively reflects the lower half of the left vertical boundary without necessitating the renormalization of the spatial reflection operator. This is expressed as
\begin{equation}
\widetilde{M}^{(n)}_{x_1x_2x_1'x_2'yy'}=\sum_i {T}^{(n-1)}_{x_1x_1'yi}{T}^{(n-1)}_{x_2x_2'y'i}.\label{eq:Mtilde}
\end{equation}
Following this adjustment, the subsequent HOTRG steps proceed by treating $\widetilde{M}^{(n)}$ as if it were $M^{(n)}$. This method provides a straightforward mechanism for embodying the geometric nuances of a rainbow state within the tensor network formalism, leveraging the inherent symmetries and structural modifications to simulate complex boundary conditions effectively.

In this construction, we get the rainbow boundary state by just contracting the indices of the largest eigenvector as shown in Fig.~\ref{fig:rainbow-reflect}:
\begin{equation}
    \widetilde{F}_{\mathcal{R}}^{(n)} = \ln \qty\Bigg(\sum_{x}\widetilde{L}_{xx}^{(n)}), \label{eq:fr-refl}
\end{equation}
where $\widetilde{L}^{(n)}$ is the largest eigenvector obtained from diagonalizing $\widetilde{M}^{(n)}$ in Eq.~\eqref{eq:Mtilde}~\footnote{One might expect that the crosscap free energy can also be calculated in this case using the spatial reflection operator, but it does not work. The reason would be that the isometry ${\widetilde{U}}^{(n)}_{xx_1x_2}$ becomes highly symmetric in the $y$ direction (to the exchange between $x_1 \leftrightarrow x_2$) and the information of reflection tends to be dropped by truncation.}.

\section{\label{sec:results}Results}

\begin{figure}[tb]
    \centering
    \includegraphics[width=\linewidth]{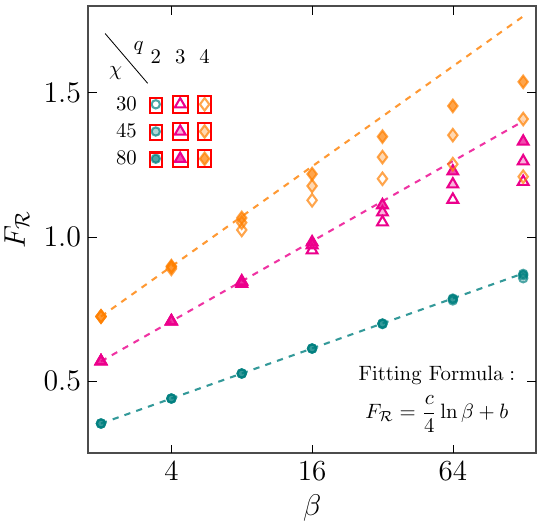}
    \caption{The rainbow free energy term $F_\mathcal{R}$ of the $q$-state Potts models ($q=2,3$ and $4$) obtained from HOTRG with the bond dimension $\chi = 30, 45$ and $80$. For each model, these values are obtained by both the renormalization of the spatial reflection operator as in Eq.~\eqref{eq:fr-sro} and by reflecting the left-hand side isometries as in Eq.~\eqref{eq:fr-refl}, which are consistent and indistinguishable in the figure. They are compared with the theoretical predictions denoted by the dashed lines.}
    \label{fig:fr}
\end{figure}

\begin{figure}[tb]
    \centering
    \includegraphics[width=\linewidth]{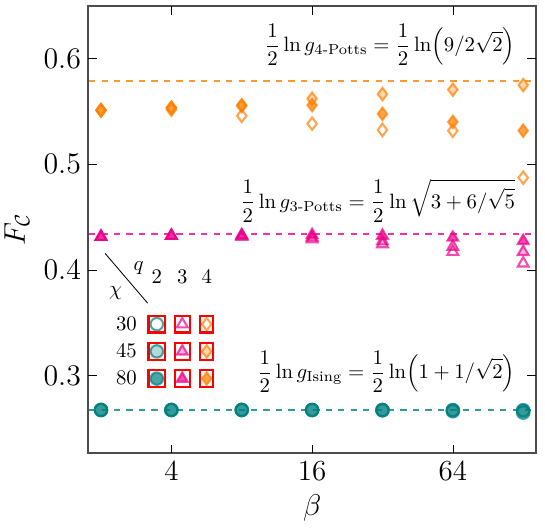}
    \caption{The crosscap free energy term $F_\mathcal{C}$ of the $q$-state Potts models ($q=2,3$ and $4$) obtained from HOTRG with the bond dimension $\chi = 30, 45$ and $80$. For each model, these values are obtained by the renormalization of the spatial reflection operator as in Eq.~\eqref{eq:crosscap}. They are compared with the theoretical predictions denoted by the dashed lines.}
    \label{fig:fc}
\end{figure}

We implemented the crosscap and rainbow boundary conditions on the $q$-state Potts models with $q=2,3$ and $4$ to calculate the free energy terms $F_\mathcal{C}$ and $F_\mathcal{R}$. Note that the $q=2$ case corresponds to the Ising model. As shown in Fig.~\ref{fig:fr}, the rainbow free energy term $F_\mathcal{R}$ exhibits excellent agreement with the theoretical scaling in Eq.~\eqref{eq:zrp2-scaling} up to approximately $\beta \sim 128$ for $q=2$, $\beta \sim 32$ for $q=3$, and $\beta \sim 16$ for $q=4$. The rainbow free energy term is also calculated via Eq.~\eqref{eq:fr-refl}, with the resulting $\widetilde{F}_\mathcal{R}$ in perfect agreement with $F_\mathcal{R}$, the value obtained by using the spatial reflection operator as in Eq.~\eqref{eq:fr-sro}. 

Regarding the crosscap free energy $F_\mathcal{C}$, as shown in Fig.~\ref{fig:fc}, the results for $q=2$ and $3$ align well with Eq.~\eqref{eq:zk-scaling}. However, the $q=4$ model ($\chi=80$) exhibits a deviation which is consistent with previous findings~\cite{tang_universal_2017}, where it is conjectured to stem from marginally irrelevant perturbations. We attribute the apparent accuracy at $\chi=45$ to a fortuitous cancellation of errors. Aside from this exception, our numerical results systematically approach theoretical predictions as the bond dimension increases. These results underscore the effectiveness and precision of our proposed methodologies.

\begin{figure}[tb]
    \centering
    \includegraphics[width=\linewidth]{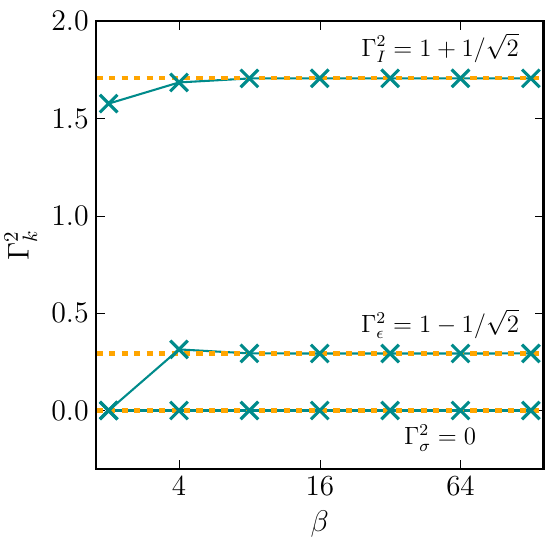}
    \caption{$\Gamma_k$ obtained from the Ising model by HOTRG with the bond dimension $\chi = 45$. The crosscap boundary state is constructed in the way described in Fig.~\ref{fig:crosscap}.}
    \label{fig:Gammak}
\end{figure}

Finally, we proceed to calculate another set of CFT data, specifically $\Gamma_k$, as outlined in Eq.~\eqref{eq:Gammak}. This equation can also be interpreted as the overlap $\langle\phi_k|\mathcal{C}\rangle$, which allows us to compute $\Gamma_k$ by evaluating the overlap between the normalized primary state $|\phi_k\rangle$ and the crosscap state $|\mathcal{C}\rangle$. Recalling that the transfer matrix on a cylinder is given by $e^{-2\pi(L_0+\bar{L}_0)}$, where $L_0$ and $\bar{L}_0$ represent the holomorphic and anti-holomorphic components of the dilatation operators, respectively, its eigenvectors correspond to the primary operators and their descendants. Importantly, the leading eigenvector $|i_0\rangle$ in the previous sections is associated with the identity operator $\mathbb{I}$. Consequently, we derive the following relation:

\begin{align*}
\Gamma_{\mathbb{I}}^2 &= |\langle 0|\mathcal{C}\rangle|^2,\\
&= g.
\end{align*}
To compute $\Gamma_k$ for a generic $k$, we utilize the methodology outlined in Sec.~\ref{sec:crosscap} and apply it to additional eigenvectors. Specifically, each eigenvector of the transfer matrix, constructed following the procedure described in Sec.~\ref{sec:crosscap}, possesses two indices. The overlap with the crosscap state is determined by contracting these indices similarly to Eq.~\eqref{eq:crosscap}. To illustrate this process, we apply our method to the Ising model. Within this context, the second leading and third leading eigenvectors correspond to the magnetic operator $\sigma$ and the energy operator $\epsilon$, respectively, in the Ising CFT. The numerical results are summarized in Fig.~\ref{fig:Gammak}.
$\Gamma_k$ of the Ising model has analytical solutions: $\Gamma_{\mathbb{I}}^2=\frac{2+\sqrt{2}}{2}$, $\Gamma_\sigma^2=0$, and $\Gamma_\epsilon^2=\frac{2-\sqrt{2}}{2}$. Our numerical results are highly consistent with the theoretical values, denoted by dotted lines. Similarly, computing the same quantities on other lattice models enables direct access to the universal information about nonorientable manifolds $\Gamma_k$, which highlights the power of our scheme.

\section{\label{sec:discussion-summary}Conclusion and Discussion}

\begin{figure}
    \centering
    \includegraphics[width=0.5\linewidth]{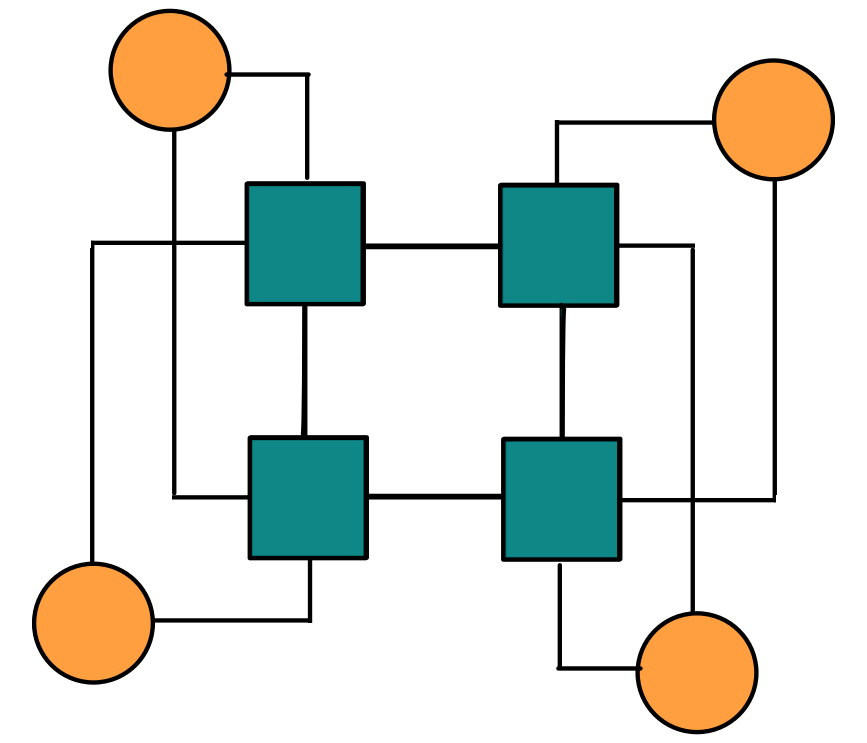}
    \caption{Diagrammatic explanation of how to construct a genus-two orientable surface. The green square tensors are the fixed point tensors obtained from HOTRG, and the orange circle tensors are the spatial reflection operators which transform the horizontal basis of the bulk tensors to the vertical basis.}
    \label{fig: g2}
\end{figure}

In our work, we introduced a method for calculating the partition functions on nonorientable surfaces, fundamentally advancing tensor network simulations. The core of our approach hinges on maintaining data on boundary configurations during the coarse-graining steps by tracking a spatial reflection operator, denoted as $\mathbf{\it O}$. While a standard leg contraction reflects periodic boundary conditions, additionally incorporating $\mathbf{\it O}$ enables the theoretical construction of partition functions across diverse manifolds. For instance, an orientable genus-$n$ surface can be realized by contracting (sewing) $2n$ renormalized tensors, and we can extend this idea to nonorientable ones by inserting $\mathbf{\it O}$ when sewing the tensors. For example, Fig.~\ref{fig: g2} shows how to construct such surfaces diagrammatically. Here, the orange circles are the spatial reflection operator, which is constructed as the transformation matrix from the basis of vertical bonds to the basis of horizontal bonds of the renormalized tensor depicted as green square tensors. We would be able to obtain the universal value by taking the thermodynamic limit with this geometry. This development not only broadens the applicability of tensor network simulations but also opens new avenues for exploring the physics of complex geometries and topologies. 
As a practical application, we showcase the calculation of the one-point function on the nonorientable surface $\mathrm{RP}^2$, addressing a previously unexplored aspect of universal information within tensor network research. By synthesizing our method with insights from prior studies, we establish a comprehensive framework capable of deriving all critical data pertinent to CFT from tensor network analyses. This achievement underscores the potential of our approach to contribute significantly to the understanding of CFT data and its implications for phase transitions and critical phenomena in statistical physics and condensed matter theory.

\section*{Acknowledgements}
A.~U.~is supported by the MERIT-WINGS Program at the University of Tokyo, the JSPS fellowship (DC1). He was supported in part by MEXT/JSPS KAKENHI Grant No. JP22KJ0565. A part of the computation in this work was done
using the facilities of the Supercomputer Center, the Institute for Solid State Physics, the University of Tokyo.

\bibliography{apssamp}

\end{document}